\documentclass[sigconf]{acmart}
\AtBeginDocument{%
  \providecommand\BibTeX{{%
    \normalfont B\kern-0.5em{\scshape i\kern-0.25em b}\kern-0.8em\TeX}}}

\setcopyright{acmcopyright}
\copyrightyear{2021}
\acmYear{2021}
\acmDOI{10.1145/1122445.1122456}

\acmConference[KDD '21]{KDD '21: SIGKDD International Conference on Knowledge Discovery and Data Mining}{August 14--18, 2021}{Singapore}
\acmBooktitle{KDD '21: SIGKDD International Conference on Knowledge Discovery and Data Mining,
  August 14--18, 2021, Singapore}
\acmPrice{15.00}
\acmISBN{978-1-4503-XXXX-X/18/06}

\begin{document}

\title{Independent Ethical Assessment of Text Classification Models: A Hate Speech Detection Case Study
}


\author{Amitoj Singh}
\authornotemark[1]
\affiliation{%
  \institution{Emerging Technology PwC}
  \city{PwC Acceleration Center}
  \state{Mumbai}
  \country{India}
}

\author{Jingshu Chen}
\authornotemark[1]
\affiliation{%
  \institution{PwC Acceleration Center Shanghai}
  \city{Pudong New District}
  \state{Shanghai}
  \country{China}}

\author{Lihao Zhang}
\authornotemark[1]
\affiliation{%
  \institution{PwC Acceleration Center Shanghai}
  \city{Pudong New District}
  \state{Shanghai}
  \country{China}
  \authornote{Authors contributed equally to this research.}
}

\author{Amin Rasekh}
\authornotemark[2]
\affiliation{%
 \institution{Product Innovation PwC}
 \city{Washington}
 \state{DC}
 \country{USA}
 \authornote{Corresponding author: amin.rasekh@pwc.com}
}

\author{Ilana Golbin}
\affiliation{%
  \institution{Emerging Technology PwC}
  \city{Los Angeles}
  \state{CA}
  \country{USA}}

\author{Anand S. Rao}
\affiliation{%
  \institution{Global AI Leader PwC}
  \city{Boston}
  \state{MA}
  \country{USA}}



\renewcommand{\shortauthors}{Amitoj and Chen, et al.}

\begin{abstract}
An independent ethical assessment of an artificial intelligence system is an impartial examination of the system's development, deployment, and use in alignment with ethical values. System-level qualitative frameworks that describe high-level requirements and component-level quantitative metrics that measure individual ethical dimensions have been developed over the past few years. However, there exists a gap between the two, which hinders the execution of independent ethical assessments in practice. This study bridges this gap and designs a holistic independent ethical assessment process for a text classification model with a special focus on the task of hate speech detection. The assessment is further augmented with protected attributes mining and counterfactual-based analysis to enhance bias assessment. It covers assessments of technical performance, data bias, embedding bias, classification bias, and interpretability. The proposed process is demonstrated through an assessment of a deep hate speech detection model.
\end{abstract}

\ccsdesc[300]{Computing methodologies~Supervised learning by classification}
\ccsdesc[300]{Computing methodologies~Natural language processing}



\keywords{natural language processing, classification, responsible AI, ethical assessment, fairness}


\maketitle

\section{Introduction}
Internal and external independent assessments of artificial intelligence (AI) systems are conducted to examine quality, responsibility and accountability for their development, deployment, and use \cite{high2020assessment, madiega2019eu}. Major ethical concerns featuring language models have recently attracted a lot of attention for the need of accountability in this domain for example: Microsoft's AI Chatbot Tay tweeted racists comments when trained on Twitter Data and GPT-3 AI generated blog hit no. 1 on Hacker News which was completely fake, hinting a mass misinformation threat. Independence in this context is defined as the freedom from conditions that may jeopardize the ability of the assessment activity to fulfill assessment responsibilities in an impartial manner \cite{institute2021international}. A three-level classification of independent assessments is provided in Appendix \ref{appendix:three-level-IAC}.

An independent ethical assessment requires both system-level qualitative guidelines that describe the process and requirements as well as component-level quantitative tools that enable the calculation of assessment metrics related to individual ethical dimensions. The Assessment List for Trustworthy Artificial Intelligence \cite{high2020assessment} and the Artificial Intelligence Auditing Framework \cite{the2017artificial} are examples of such qualitative guidelines, amongst others \cite{information2018auditing, raji2020closing, information2020guidance}. Quantitative models that focus on different ethical components have also been developed, such as the techniques for bias detection in embedding models by \cite{papakyriakopoulos2020bias, garg2018word, rozado2020wide} and the metrics for evaluating bias in text classification datasets and models by \cite{dixon2018measuring, blodgett2020language, hutchinson2020social, park2018reducing, sap2019risk}. Quantitative measures require desired metrics, which must be defined by the context, not by the framework. 

There exists a gap between system-level qualitative guidelines and component-level quantitative models that hinders execution of independent ethical assessments in practice \cite{brown2021algorithm}. This study bridges this gap for text classification systems in the context of hate speech detection models. While hate speech detection aims to serve the positive cause of reducing hateful content, it also poses the risk of silencing harmless content, or even worse, doing so with a false positive bias and without an ability to provide an explanation of its reasoning.

We summarize our main contributions as follows:
\begin{itemize}
    \item We develop a process for independent ethical assessments of text classification systems by bridging high-level qualitative guidelines and low-level technical models.
    \item We propose methods to mine protected attributes from unstructured data.
    \item We demonstrate approaches and metrics for quantifying bias in data, word embeddings, and classification models.
\end{itemize}

The paper is organized as follows. Section \ref{section:process-desc} presents the proposed assessment process. Section \ref{section:sys-tech-assess} describes the text classification system under consideration and its technical assessment. Section \ref{section:data-bias-assess} presents the data bias assessment using two different approaches. Section \ref{section:embed-bias-assess} describes the embedding bias metrics and assessment. Section \ref{section:classify-bias-assess} presents the classification bias assessment using the original, swapped, and synthetic datasets. Section \ref{section:classify-interpret-assess} describes the local and global interpretability assessment. Section \ref{section:conclusion} concludes the study.

\section{Process Description}
\label{section:process-desc}
This process performs quantitative assessments to address the qualitative requirements set by the Assessment List for Trustworthy Artificial Intelligence (ALTAI) by the European Commission \cite{high2020assessment}. The classification accuracy assessment addresses Requirement \#2 Technical Robustness and Safety. Bias assessments are conducted for the data, word embeddings, and classification model, which correspond to Requirement \#5 Diversity, Non-discrimination and Fairness. ALTAI defines bias as “systematic and repeatable errors in a computer system that create unfair outcomes, such as favoring one arbitrary group of users over others.” Data and embedding bias are possible causes of unfairness, and may influence classification bias, which is an actual measure of fairness.  Protected attributes are defined as those qualities, traits, or characteristics, such as gender, that cannot be discriminated against. Model interpretability assessments are performed to address Requirement \#4 Transparency. Since the AI system under consideration is not computationally intensive and was trained on a laptop, its environmental impact is deemed to be negligible. Nevertheless, Requirement \#6 Societal and Environmental Well-being is briefly addressed in Appendix \ref{appendix:environ-impact-assess}. Indeed, this process does not cover all requirements and future work is necessary to address the remainder.


\section{The AI System and Technical Assessment}
\label{section:sys-tech-assess}
The system under assessment is a pre-trained natural language processing (NLP) system that identifies hateful comments on a social media platform. The system has been trained on a dataset of Twitter posts in English and classifies comments/posts as hateful or non-hateful. It is a proof of concept model that was developed by a separate team within the same organization. Information about the dataset and model are provided in Appendix \ref{appendix:data-n-model-desc}.

Requirement \#2 in ALTAI states that reliable technical performance of the system is one of the critical requirements for trustworthy AI. We assessed the performance of the model on the test dataset using various metrics. The results are presented in Table \ref{tab:tech-perf-assess} and the accuracy of the model is 88\%.

\begin{table}[ht]
  \caption{Technical performance assessment of the AI system. Macro and weighted averages are calculated through averaging the unweighted mean and the support-weighted mean per label, respectively.}
  \label{tab:tech-perf-assess}
  \begin{tabular}{ccccc}
    \toprule
    &Precision&Recall&F1-Score&Support\\
    \midrule
    Not-hateful & 0.78& 0.92& 0.84& 1,031\\
    Hateful & 0.95& 0.87& 0.91& 2,502\\
    Macro Avg. & 0.87& 0.89& 0.87& 3,803\\
    Weighted Avg. & 0.89& 0.88& 0.88& 3,803\\
   \bottomrule
  \end{tabular}
\end{table}


\section{Data Bias Assessment}
\label{section:data-bias-assess}
Requirement \#5 in ALTAI outlines the need for an assessment of the input data for bias. This data bias assessment step is motivated by the possible impact that data bias may have on the fairness of the classification model. In the case of structured datasets, correlations between protected attribute features and decision variables can be analyzed to assess data bias. For unstructured data like text, however, this approach may not be possible because protected attributes are not represented as separate, standalone features. This study uses two approaches to tackle this problem and assess bias in the training data.

In the first approach, which is inspired by \cite{dixon2018measuring}, a curated set of identity terms are first identified and their frequencies across hateful and all the comments in the training dataset are then found and compared. The results for all the identity terms are provided in Appendix \ref{appendix:freq-identity-terms}. They indicate that some of identity terms such as “gay” and “white” are disproportionately used in hateful comments and this could be an indicative of false positive bias.

The second approach is based on the frequency of protected attribute references across hateful comments and overall. Protected attribute mining is first conducted to identify references made to each subgroup of each attribute in the comments (as elaborated upon in Section \ref{section:bias-assess-protected-attr} below). Their frequencies across the hateful comments and all the comments is then calculated and compared. The results are shown in Table \ref{tab:protected-attr-freq} for the protected attributes of gender and religion. The data shows some bias is present in the data such as for the “Islam” subgroup and the "female" subgroup. The word Islam, in our dataset, has a larger presence in the hateful comments as compared word Christianity. Same holds true for the word female compared to the word male.

\begin{table}[ht]
  \caption{Frequency of protected attribute references in hateful and not-hateful comments and overall for the protected attributes of gender and religion}
  \label{tab:protected-attr-freq}
  \begin{tabular}{p{0.14\linewidth}cp{0.12\linewidth}cp{0.13\linewidth}}
    \toprule
    \centering Protected Attribute&Subgroup&Hateful&Not-hateful&Overall\\
    \midrule
    \centering Gender & Female & 12.28\% & 11.53\% & 12.02\% \\
    & Male & 11.51\% & 17.37\% & 13.52\% \\ \hline
    \centering Religion & Islam & 0.10\% & 0.15\% & 0.12\% \\
    & Christianity & 0.07\% & 0.24\% & 0.13\% \\
   \bottomrule
  \end{tabular}
\end{table}


\section{Embedding Bias Assessment}
\label{section:embed-bias-assess}
The embedding model should be assessed for bias as it influences the classification model's decisions, which relates to Requirement \#5 in ALTAI. This embedding bias assessment is motivated by the potential influence that embedding bias has on the fairness of the classification model. To assess embedding bias, we analyze similarity between a diverse set of 501 neutral words (e.g., “admirable” and “miserable”) \cite{garg2018word} and word groups associated with three protected attributes of gender, religion, and ethnicity (Appendix \ref{appendix:protected-attr-word-group}). Neutral words mean they are expected to be similarly correlated across protected attribute words \cite{garg2018word}. The similarity metric used is cosine similarity between the individual terms' embeddings and is aggregated using average mean absolute error (AMAE) and average root mean squared error (ARMSE).

\begin{figure}[h]
  \centering
  \includegraphics[width=0.5\linewidth,height=2.7cm]{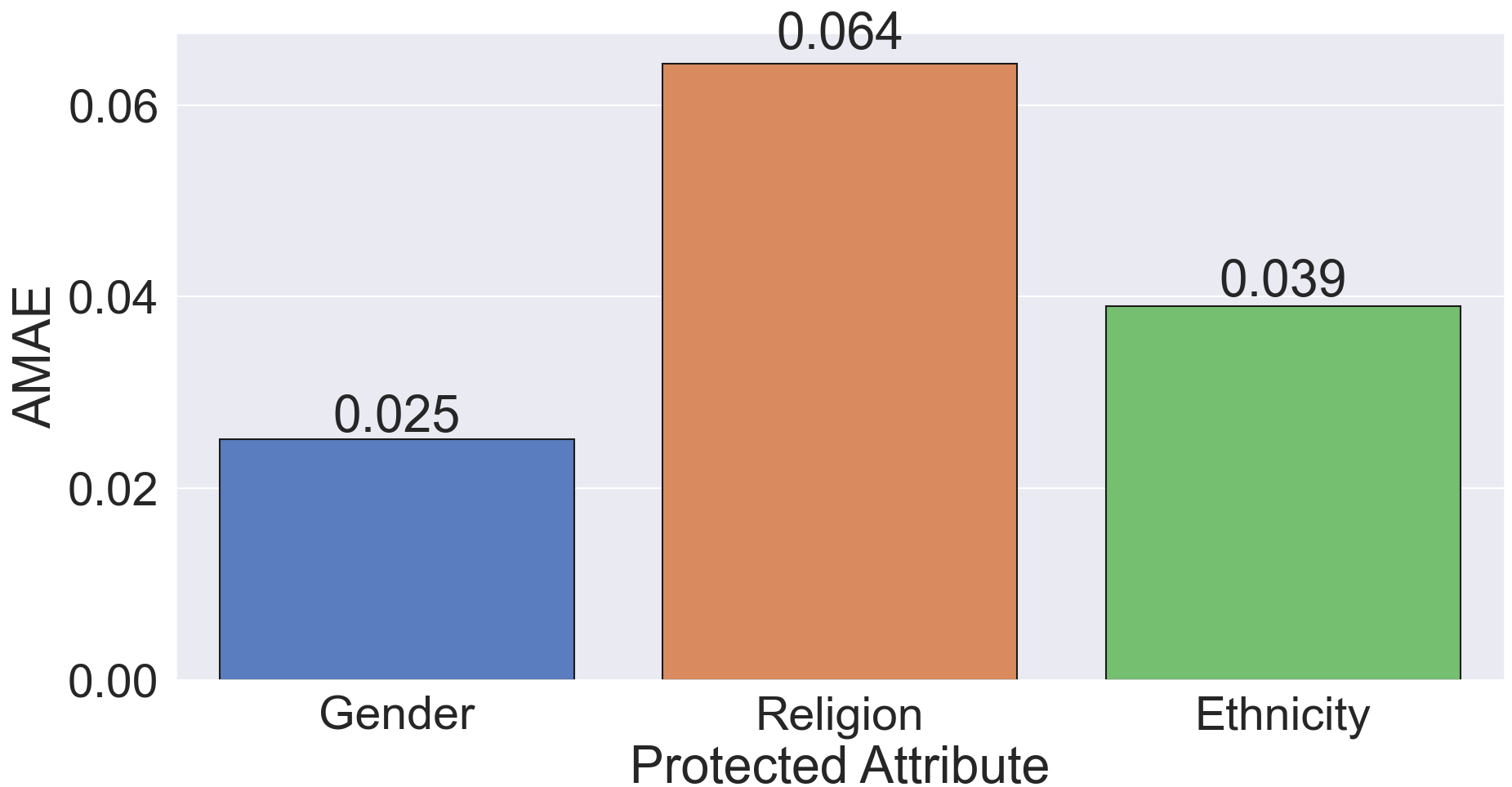}\hfill
  \includegraphics[width=0.5\linewidth,height=2.7cm]{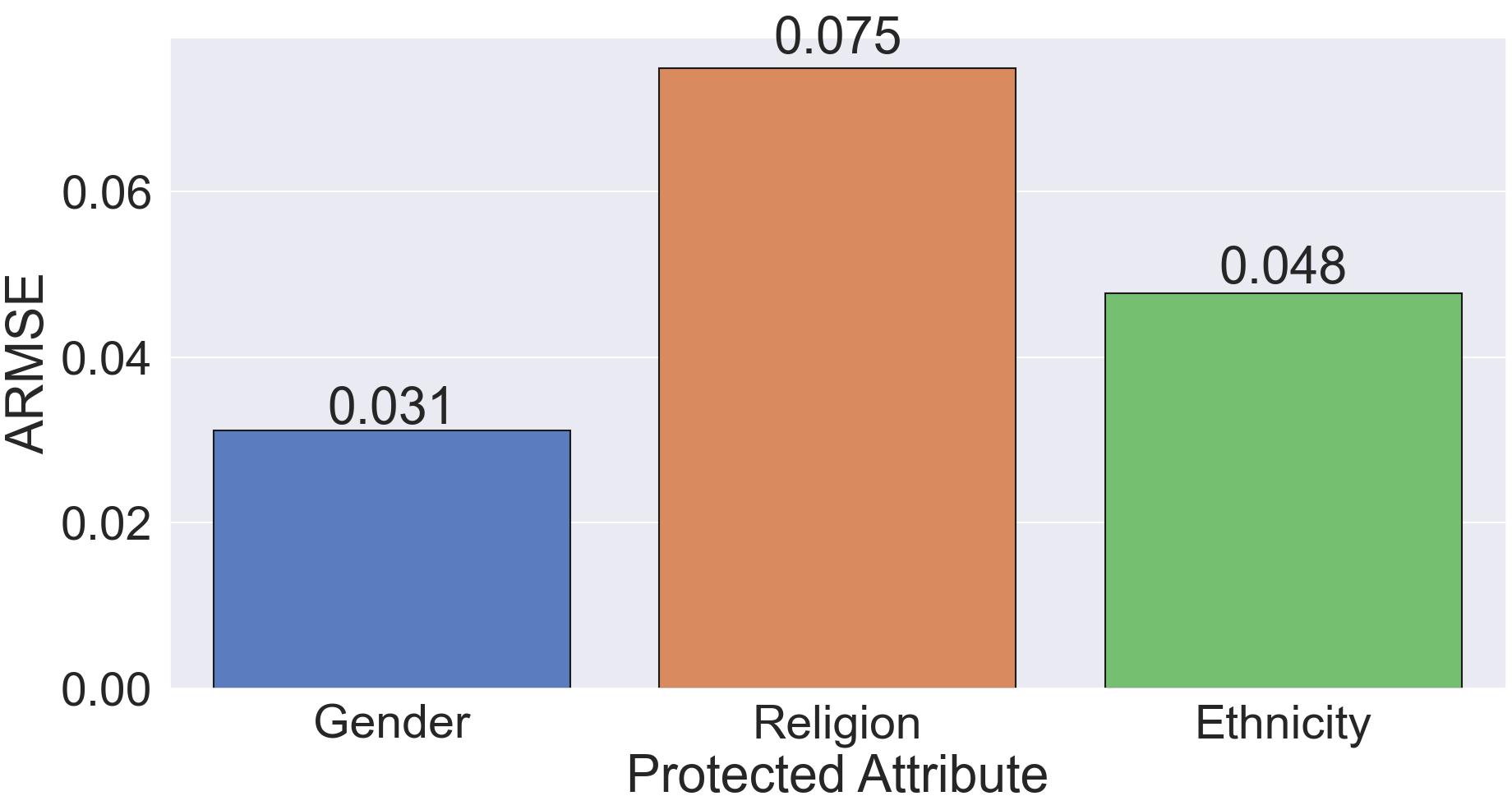}\hfill
  \caption{Embedding bias magnitude across different protected attributes}
  \label{fig:embedding-bias}
\end{figure}

Given a protected attribute (e.g., gender) and a subgroup within (e.g., female), the similarity between an individual neutral word and that subgroup is calculated by averaging distances for the individual neutral word and subgroup word, such as, for example, “nice” and “she”. This is done for all the neutral words and subgroups. Then the AMAE and ARMSE of the average cosine similarity across the subgroups is calculated (the equations are provided in Appendix \ref{appendix:embed-bias-metric-n-val}). This is performed for all the protected attributes and the results are illustrated in Figure \ref{fig:embedding-bias}. The magnitudes of the AMAE and ARMSE show how strong the biases are with that particular category. In principal, the 2 measures are similar to the statistical measure of "variance" of cosine-distances across the various sub-groups in a category. Therefore, higher the measure, greater is the disparity in association and more biased are the embeddings for that category.

The AMAE and ARMSE scores indicate that the embedding model is more biased as it relates to religion terms but less for gender terms. Appendix \ref{appendix:embed-bias-metric-n-val} provides a comparison of the bias magnitude for this system's embedding model against other mainstream embedding models.


\section{Classification Bias Assessment}
\label{section:classify-bias-assess}
While classification accuracy at an aggregate level is related to ALTAI Requirement \#2 Technical Robustness and Safety, an accuracy report broken down by various protected attributes would address Requirement \#5 Diversity, Non-discrimination and Fairness. In contrast to data and embedding bias, classification bias may directly impact the system's stakeholders.

To assess this bias, subgroups based on gender, ethnicity, religion, etc. are first identified in different comments. Bias assessment is then performed to understand how the model’s prediction accuracy differs across these protected attributes. This assessment can be enhanced by measuring changes in the model's accuracy as the subgroups in the comments are swapped (for example, does the model's true positive prediction for "men are universally terrible" changes to a false negative if "men" is swapped with "women"?). Synthetic sentences containing protected attribute references are also generated and analyzed for a better understanding of how the model learns biases.

\subsection{Protected Attributes Mining}
To extract protected attributes from text, two approaches are applied in this study: 1) a look-up based approach and 2) a named-entity recognition (NER) based approach. These approaches have been validated using datasets with human-annotated comments \cite{dixon2018measuring}. The look-up approach is based on the word counts extracted from the text examples and the words are looked up in a list of common identity terms such as pronouns and professions. The NER based approach uses spaCy’s pre-trained NER model to extract entities (i.e. protected attributes). A named entity is a “real-world object” with a type associated with it, which might be a country, a product, a religion, etc. In this study, entities tagged with the “NORP” type are extracted. “NORP” refers to the “nationalities or religious or political groups”. spaCy’s NER model takes into account the context and part of speech to determine the tag for a given word. Gender and religion are the protected attributes addressed in this subsection. Several examples of protected attribute extractions from comments are included in Appendix \ref{appendix:classify-bias-supplements}.


\subsection{Bias Assessment based on Protected Attributes Mining}
\label{section:bias-assess-protected-attr}
\begin{figure}[h]
  \centering
  \includegraphics[width=0.9\linewidth]{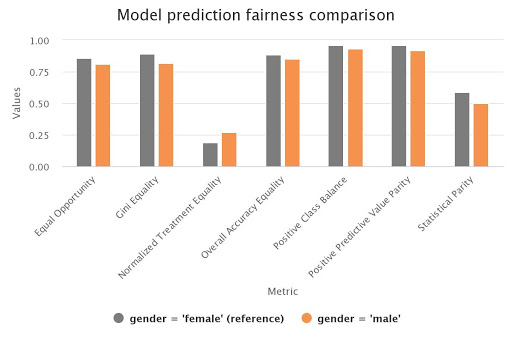}\hfill
  \bigbreak
  \includegraphics[width=0.9\linewidth]{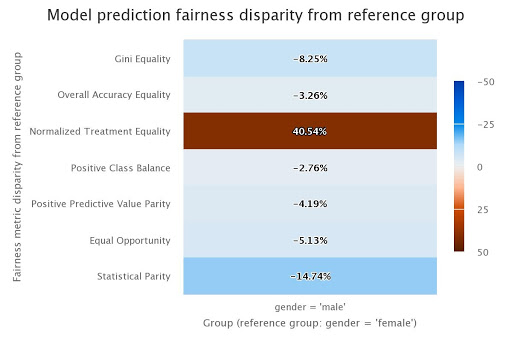}\hfill
  \caption{Bias assessment based on protected attribute extraction for gender}
  \label{fig:bias-assess-gender}
\end{figure}

The PwC Responsible AI (RAI) Toolkit performs a bias assessment using protected attribute references, ground-truth labels, and prediction probabilities \cite{rao2019practical}. It checks whether the model discriminates against various groups or individuals using a variety of bias metrics. Table \ref{tab:bias-assess-gender} and Figure \ref{fig:bias-assess-gender} show the results for gender. There were 501 and 442 references to male and female subgroups, respectively. Appendix \ref{appendix:bais-metric-def} provides a description of the seven bias metrics used.

\begin{table}[ht]
  \caption{Average prediction probabilities for comments with gender references}
  \label{tab:bias-assess-gender}
  \begin{tabular}{ccc}
    \toprule
    Actual Comment Type & Subgroup & Avg. Predicted Probability \\
    \midrule
    Not-hateful & Female & 14.0\% hateful \\
    Not-hateful & Male & 15.4\% hateful \\
    Hateful & Female & 85.6\% hateful \\
    Hateful & Male & 80.4\% hateful \\
   \bottomrule
  \end{tabular}
\end{table}

\subsection{Assessment Based on Swapped Protected Attributes}
Versions of the comments where identity terms are swapped help address the limitations of relying solely on the ground-truth data such as in the case of disproportionate representation of one subgroup over another. Examples are provided in \ref{appendix:classify-bias-supplements}. 




A bias assessment can be performed using the actual label and the predicted probabilities before and after identity swapping. We define the term 'favor' as the model's confidence in its prediction one way or the other. For not-hateful comments, a decrease in the prediction probability after swapping identities (e.g. male to female) means the model favors the new identity. For hateful comments, an increase in the prediction probability after swapping identities (e.g. male to female) means the model favors the new identity. This process of swapping is repeated for all the comments with the protected attribute references. This bias assessment indicated that for 55.9\% of the comments the model favored the female subgroup, while in 43.4\% of the comments it favored the male subgroup. The model prediction did not change for 0.7\% of the comments with probabilities rounded to four decimal places.

\subsection{Bias Assessment Based on Synthetic Counterfactual Dataset}
Bias assessment can also be performed using a synthetic counterfactual dataset. This method offers greater flexibility but is labor-intensive and the resulting dataset may not be representative of the actual texts the model encounters in practice. This idea is related to counterfactual fairness, which as stated by \cite{kusner2017counterfactual}, “captures the intuition that a decision is fair towards an individual if it is the same in (a) the actual world and (b) a counterfactual world where the individual belonged to a different demographic group”. Examples of such counterfactuals for religion are provided in Appendix \ref{appendix:classify-bias-supplements}. 

Once the synthetic counterfactual dataset is generated, a bias assessment, similar to the above swapped dataset assessment, can be conducted. We used several templates (some inspired from \cite{dixon2018measuring}) for gender and religion. The average prediction probabilities (Appendix \ref{appendix:classify-bias-supplements}), indicate that there appears to be a bias towards treating all talk of Islam as more hateful compared to Christianity.

We use the following threshold-insensitive counterfactual bias metric for each reference subgroup:

\begin{equation}
    CB(\mathbf{X}, O, \mathbf{y}) = \sum_{i=1}^{|\mathbf{X}|} (\hat{p}_{\mathbf{x_i}} - \frac{1}{|O|}\sum^{|O|}_{j=1}\hat{p}_{O_{ij}}) \times g(y_i)
\end{equation}

\noindent where $\mathbf{X}$ is a series of reference group (e.g. female) examples (e.g. ("she is kind", "women are universally terrible", $\ldots$). $O$ is a series of sets of counterfactual examples of the form (("he is kind", $\ldots$), ("men are universally terrible", $\ldots$)). $\hat{p}_{\mathbf{x}_i} = f(\mathbf{x}_i)$ (i.e. the model's predicted probability). $\mathbf{y}$ is a series of labels for the reference group examples, where 0s correspond to not-hateful comments and 1s corresponds to hateful comments. $g(y_i)$ is -1 when $y_i = 0$ and 1 when $y_i = 1$ and is a term that corrects for the direction of the bias. For the model under consideration, the preferred label is not-hateful and therefore $g(y)$ is -1 for not-hateful comments. A positive value of $CB$ means the model favors the reference subgroup.

Setting “Islam” as the reference subgroup returns a bias value of -0.0067, which denotes the model slightly favors the “Christianity” subgroup. For gender, setting “Male” as the reference subgroup would return a bias value of 0.0737. A comprehensive assessment may also include other bias metrics such as equalized odds \cite{hardt2016equality}.



\section{Classification Interpretability Assessment}
\label{section:classify-interpret-assess}
Requirement \#4 in ALTAI focuses on transparency requirements and describes explainability as one of the elements to consider alongside traceability and open communication. The question is simple - why are certain tweets being classified as hate-speech? Understanding hate in speech is a function of the vocabulary and the context of the language used and we as humans have an innate ability to understand hate. We seek to understand how an AI model infers it. Does it consider the vocabulary? If yes, how does it do so since we have not explicitly fed the model a list of abusive words or profane phrases. Does it understand the context? If yes, how does it distinguish between cases where strong words can be used as a way of free-speech or expression (e.g. \emph{“I am Gay”}) versus cases where the words are used disparagingly as hateful comments directed at someone? We explored two methods to address these questions. Overall, the two explainability packages explored offer end-users a unique perspective to tie model decisions with the vocabulary used in a particular tweet. This may help automate, at least partially, some of the content modulation tasks Social Media players are trying to do and at the same time, provide powerful measurable evidence for driving hate. 

\begin{figure}[h]
  \centering
  \includegraphics[width=\linewidth]{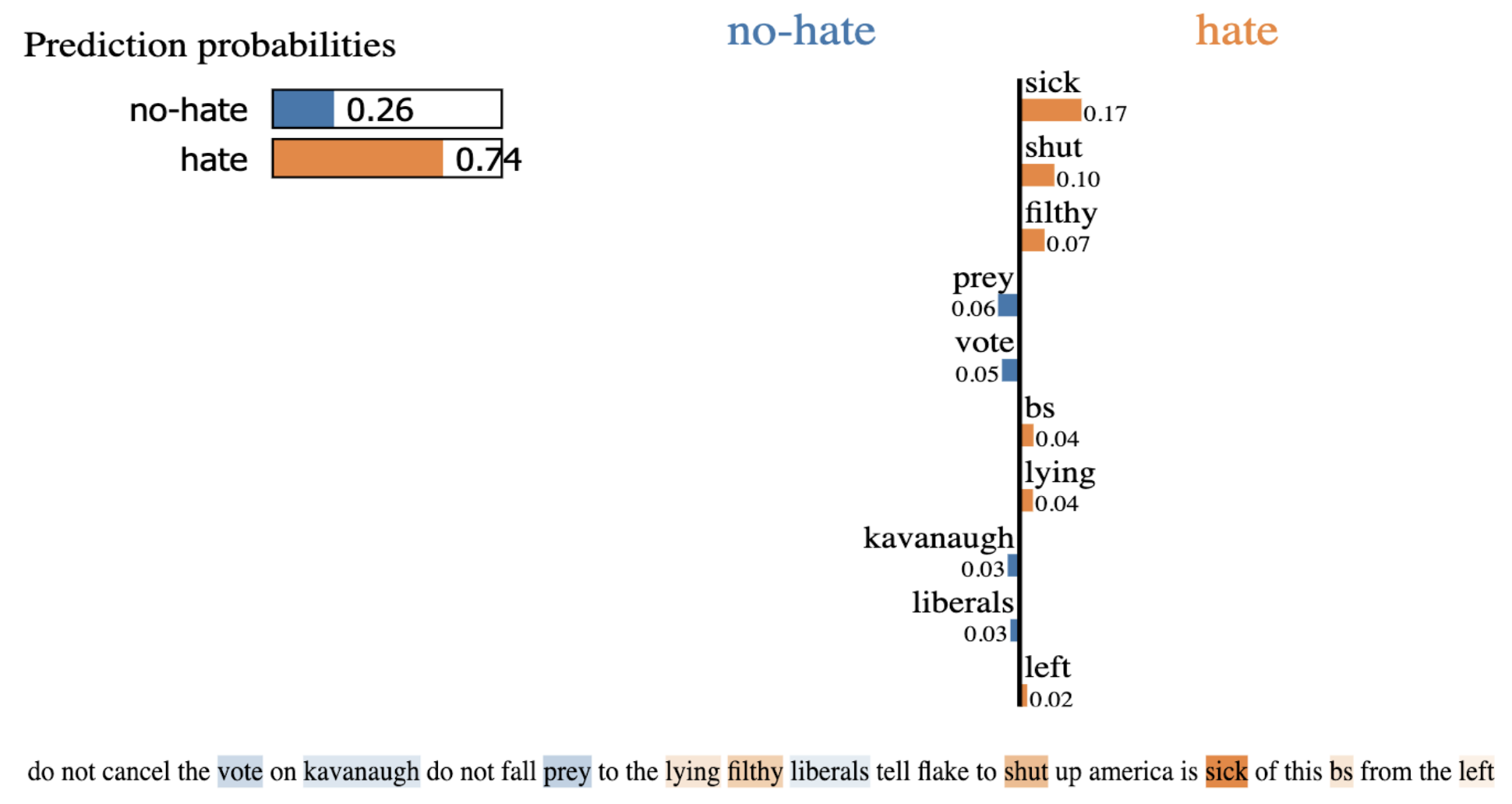}
  \caption{Local interpretability of an example prediction using LIME}
  \label{fig:lime-example}
\end{figure}

The first method focuses on local interpretability and uses LIME, which is a model agnostic method that helps explain a given prediction through the features used in the model \cite{ribeiro2016should}. In the case of a text classification model, LIME assigns each token in a given sentence an importance score which can either be positive or negative, depending on the word’s effect on the prediction. For example, in the tweet shown in Figure \ref{fig:lime-example}, LIME shows that words like “vote” and “liberals” contribute to the model predicting non-hate, while words like “filthy” and “sick” push the model to classify as a hate-comment. LIME can be used to look at incidents of hate tweets as shown in Fig 3. to understand how the AI model determines hate. This can be of important use for the en-user to better analyze the classification and what kind of vocabulary is too strong for the model to predict the comment as non-hateful. Here the use of strong-language was the key for the model to predict hate.

The second method concerns global interpretability and uses SHAP values. SHAP is a model-agnostic explainability tool that generates feature importances using the overall datasets. It produces various permutations over feature selection to get an estimate of the importance of a particular feature on the overall model \cite{lundberg2017unified}. The global interpretability results are provided in Appendix \ref{appendix:global-interpret-result}. The results show at an overall level that strong-language is what drives model in predicting hate. The keywords listed here had the largest impact on model's detection of hate-speech. 



\section{Conclusions}
\label{section:conclusion}
This study proposed a process for independent ethical assessments of text classification systems and demonstrated it through a first-party assessment of a hate speech detection model. While the ALTAI recommendations are largely applicable to text classification systems, they are underspecified, leaving a looming implementation gap. This study addresses that gap through a series of component-wise quantitative assessments. A discussion of the other requirements with ideas for addressing them are provided in the Appendix \ref{appendix:environ-impact-assess}. Risk mitigation is also a logical next step to any risk assessment process. This work serves as a concrete blueprint for the design and implementation of independent ethical assessment processes for text classification systems. More work is needed to apply these ideas to other domains.

\begin{acks}
We wish to thank Todd Morrill for thorough feedback on this article.
\end{acks}

\bibliographystyle{ACM-Reference-Format}
\bibliography{sample-base}

\appendix

\section{A three-level independent assessment classification}
\label{appendix:three-level-IAC}
The three-level classification that is common in the audit and assurance landscape \cite{russell2019asq} can be adapted. 

A first-party ethical assessment of an AI system is an internal assessment conducted by a team within the same organization that owns the AI system but who had no role in its development nor has any vested interest in its assessment results. 

A second-party assessment is an external assessment conducted by a contracted party outside the organization on behalf of a customer. 

A third-party assessment is an external assessment by a party independent of the organization-customer relationship and assesses the level of conformity of the AI system to certain ethical criteria and standards \cite{russell2019asq}.

The AI system development team and the AI system assessment team are the two primary groups involved in the process for a first-party ethical assessment. The two teams are independent of each other, although they are still within the same organization. The AI system development team is analogous to the first and second lines in the IIA’s Three Line Model \cite{institute2021international}, which are the creators, executors, operators, managers, and supervisors who are in charge of designing, building, and deploying the AI system as well as risk assessment and strategy. The AI system assessment team is related closely to the third line, which are the auditors and ethicists who assess the system against the ethical requirements.

\section{Dataset and model description}
\label{appendix:data-n-model-desc}
\emph{The text dataset:} The dataset has been created by combining two public datasets from \cite{davidson2017automated, zampieri2019predicting} and used for the purpose of training the hate speech classifier. The model was trained on 34,220 Tweets and an unseen test dataset of 3803 Tweets was also provided to us.\\

\noindent\emph{The word embedding model:} The classifier uses the pretrained Stanford Glove Common Crawl word embeddings with 300 dimensions \cite{pennington2014glove} for training and predictions.\\

\noindent\emph{The classification model:} The classifier is a convolutional neural network model that takes word embeddings as input. The network architecture is a 1D CNN (32 filters with a kernel size of 17), followed by a max pooling (pool size =4) layer and then two fully connected dense Layers (25 unit output followed by a 1 unit output). The assessment here uses the pretrained model as is (i.e., only to make predictions where needed) and the whole process is repeatable regardless of knowing the architecture.

\section{Frequency of identity terms in data}
\label{appendix:freq-identity-terms}
The statistics are based on the training dataset and are provided in Table \ref{tab:identity-term-freq}. It contains three columns: (1) the \% of comments labelled as “Hateful” that contain an identity term, (2) the \% of comments labelled as “Not-hateful” that contain an identity term, and (3) the \% of comments Overall (“Hateful” and “Not-hateful”) that contain an identity term.

\begin{table}[ht]
  \caption{Frequency of identity terms in hateful and not-hateful comments and overall}
  \label{tab:identity-term-freq}
  \begin{tabular}{cccc}
    \toprule
    Term&Hateful \%&Not-hateful \%&Overall \%\\
    \midrule
    atheist& 0.0044& 0.0085& 0.0058\\
    queer& 0.2220& 0.0427& 0.1607\\
    gay& 0.3864& 0.1282&  0.2981\\
    transgender& 0.0044& 0.0085& 0.0058\\
    lesbian& 0.0266& 0.0256& 0.0263\\
    homosexual& 0.0133& 0.0085& 0.0117\\
    feminist& 0.0222& 0.0000& 0.0146\\
    black& 0.6706& 0.5042& 0.6137\\
    white& 1.3367& 0.5640& 1.0725\\
    heterosexual& 0.0044& 0.0000& 0.0029\\
    islam& 0.0044& 0.0000& 0.0029\\
    muslim& 0.0178& 0.0171& 0.0175\\
   \bottomrule
  \end{tabular}
\end{table}


\section{Protected attribute word groups}
\label{appendix:protected-attr-word-group}
\subsection{Word groups for three protected attributes (based on \cite{garg2018word, zhao2018gender})}

\textbf{Religion:}\\
\indent\textbf{\emph{Islam:}} 'allah', 'ramadan', 'turban', 'emir', 'salaam', 'sunni', 'koran', 'imam', 'sultan', 'prophet', 'veil', 'ayatollah', 'shiite', 'mosque', 'islam', 'sheik', 'muslim', 'muhammad';\\
\indent\textbf{\emph{Christianity:}} 'baptism', 'messiah', 'catholicism', 'resurrection', 'christianity', 'salvation', 'protestant', 'gospel', 'trinity', 'jesus', 'christ', 'christian', 'cross', 'catholic', 'church', 'christians', 'catholics'\\

\noindent\textbf{Gender:}\\
\indent\textbf{\emph{Male:}} 'cowboy', 'cowboys', 'cameramen', 'cameraman', 'busboy', 'busboys', 'bellboy', 'bellboys', 'barman', 'barmen', 'tailor', 'tailors', 'prince', 'princes', 'governor', 'governors', 'adultor', 'adultors', 'god', 'gods', 'host', 'hosts', 'abbot', 'abbots', 'actor', 'actors', 'bachelor', 'bachelors', 'baron', 'barons', 'beau', 'beaus', 'bridegroom', 'bridegrooms', 'brother', 'brothers', 'duke', 'dukes', 'emperor', 'emperors', 'enchanter', 'father', 'fathers', 'fiance', 'fiances', 'priest', 'priests', 'gentleman', 'gentlemen', 'grandfather', 'grandfathers', 'headmaster', 'headmasters', 'hero', 'heros', 'lad', 'lads', 'landlord', 'landlords', 'male', 'males', 'man', 'men', 'manservant', 'manservants', 'marquis', 'masseur', 'masseurs', 'master', 'masters', 'monk', 'monks', 'nephew', 'nephews', 'priest', 'priests', 'sorcerer', 'sorcerers', 'stepfather', 'stepfathers', 'stepson', 'stepsons', 'steward', 'stewards', 'uncle', 'uncles', 'waiter', 'waiters', 'widower', 'widowers', 'wizard', 'wizards', 'airman', 'airmen', 'boy', 'boys', 'groom', 'grooms', 'businessman', 'businessmen', 'chairman', 'chairmen', 'dude', 'dudes', 'dad', 'dads', 'daddy', 'daddies', 'son', 'sons', 'guy', 'guys', 'grandson', 'grandsons', 'guy', 'guys', 'he', 'himself', 'him', 'his', 'husband', 'husbands', 'king', 'kings', 'lord', 'lords', 'sir', 'sir', 'mr.', 'mr.', 'policeman', 'spokesman', 'spokesmen';\\
\indent\textbf{\emph{Female:}} 'cowgirl', 'cowgirls', 'camerawomen', 'camerawoman', 'busgirl', 'busgirls', 'bellgirl', 'bellgirls', 'barwoman', 'barwomen', 'seamstress', 'seamstress', 'princess', 'princesses', 'governess', 'governesses', 'adultress', 'adultresses', 'godess', 'godesses', 'hostess', 'hostesses', 'abbess', 'abbesses', 'actress', 'actresses', 'spinster', 'spinsters', 'baroness', 'barnoesses', 'belle', 'belles', 'bride', 'brides', 'sister', 'sisters', 'duchess', 'duchesses', 'empress', 'empresses', 'enchantress', 'mother', 'mothers', 'fiancee', 'fiancees', 'nun', 'nuns', 'lady', 'ladies', 'grandmother', 'grandmothers', 'headmistress', 'headmistresses', 'heroine', 'heroines', 'lass', 'lasses', 'landlady', 'landladies', 'female', 'females', 'woman', 'women', 'maidservant', 'maidservants', 'marchioness', 'masseuse', 'masseuses', 'mistress', 'mistresses', 'nun', 'nuns', 'niece', 'nieces', 'priestess', 'priestesses', 'sorceress', 'sorceresses', 'stepmother', 'stepmothers', 'stepdaughter', 'stepdaughters', 'stewardess', 'stewardesses', 'aunt', 'aunts', 'waitress', 'waitresses', 'widow', 'widows', 'witch', 'witches', 'airwoman', 'airwomen', 'girl', 'girls', 'bride', 'brides', 'businesswoman', 'businesswomen', 'chairwoman', 'chairwomen', 'chick', 'chicks', 'mom', 'moms', 'mommy', 'mommies', 'daughter', 'daughters', 'gal', 'gals', 'granddaughter', 'granddaughters', 'girl', 'girls', 'she', 'herself', 'her', 'her', 'wife', 'wives', 'queen', 'queens', 'lady', 'ladies', "ma'am", 'miss', 'mrs.', 'ms.', 'policewoman', 'spokeswoman', 'spokeswomen'\\

\noindent\textbf{Ethnicity:}\\
\indent\textbf{\emph{Chinese:}} 'chung', 'liu', 'wong', 'huang', 'ng', 'hu', 'chu', 'chen', 'lin', 'liang', 'wang', 'wu', 'yang', 'tang', 'chang', 'hong', 'li';\\
\indent\textbf{\emph{Hispanic:}} 'ruiz', 'alvarez', 'vargas', 'castillo', 'gomez', 'soto', 'gonzalez', 'sanchez', 'rivera', 'mendoza', 'martinez', 'torres', 'rodriguez', 'perez', 'lopez', 'medina', 'diaz', 'garcia', 'castro', 'cruz';\\
\indent\textbf{\emph{White:}} 'harris', 'nelson', 'robinson', 'thompson', 'moore', 'wright', 'anderson', 'clark', 'jackson', 'taylor', 'scott', 'davis', 'allen', 'adams', 'lewis', 'williams', 'jones', 'wilson', 'martin', 'johnson'

\section{Embedding bias metrics and values}
\label{appendix:embed-bias-metric-n-val}
Let $\mathbf{M} \in \mathbb{R}^{m\times d}$ be a matrix representing $d$ dimensional word embeddings for the $m$ neutral terms (e.g. crazy, nice, etc.). Let $\mathbf{T}_i \in \mathbb{R}^{k_i\times d}$ be a matrix that corresponds to the $d$ dimensional word embeddings for the $k_i$ terms that represent subgroup $i$ (e.g. $k_1$ = 2, where 1 represents the male subgroup, which has terms \{he, boy\}). Let $\mathbf{x}_i \in \mathbb{R}^{m}$ be the vector representing the averaged cosine similarity scores between the $m$ neutral terms and the $k_i$ subgroup terms for subgroup $i$. We compute the $j^{th}$ entry of $\mathbf{x}_i$ as follows.

\begin{equation}
    x_{ij} = \frac{1}{k_i}\sum^{k_i}_{\ell=1}COSINESIM(m_j, \mathbf{T}_{i\ell})
\end{equation}

\noindent We measure deviations between subgroups (e.g. male vs female) as proxies for word embedding bias. These measures include mean absolute error (MAE), root mean squared error (RMSE), and several variations. When there are two subgroups being compared we compute

\begin{equation}
    MAE = \frac{1}{m}\sum^{m}_{j=1}|x_{1j} - x_{2j}|
\end{equation}

\noindent In the event that there are more than two subgroups being compared, we compute the average MAE (AMAE) score across all pairwise comparisons between subgroups.

\begin{equation}
    AMAE = \frac{1}{{s \choose 2}}\sum^{s}_{k=1}\sum^{s}_{n=k+1}MAE
\end{equation}

\noindent where $s$ is the number of subgroups. Similarly, we compute the RMSE and averaged RMSE (ARMSE) as follows.

\begin{equation}
    RMSE = \sqrt{\frac{\sum^{m}_{j=1}(x_{1j} - x_{2j})^2}{m}}
\end{equation}

\begin{equation}
    ARMSE = \frac{1}{{s \choose 2}}\sum^{s}_{k=1}\sum^{s}_{n=k+1}RMSE
\end{equation}

\begin{figure}[h]
  \centering
  \includegraphics[width=\linewidth]{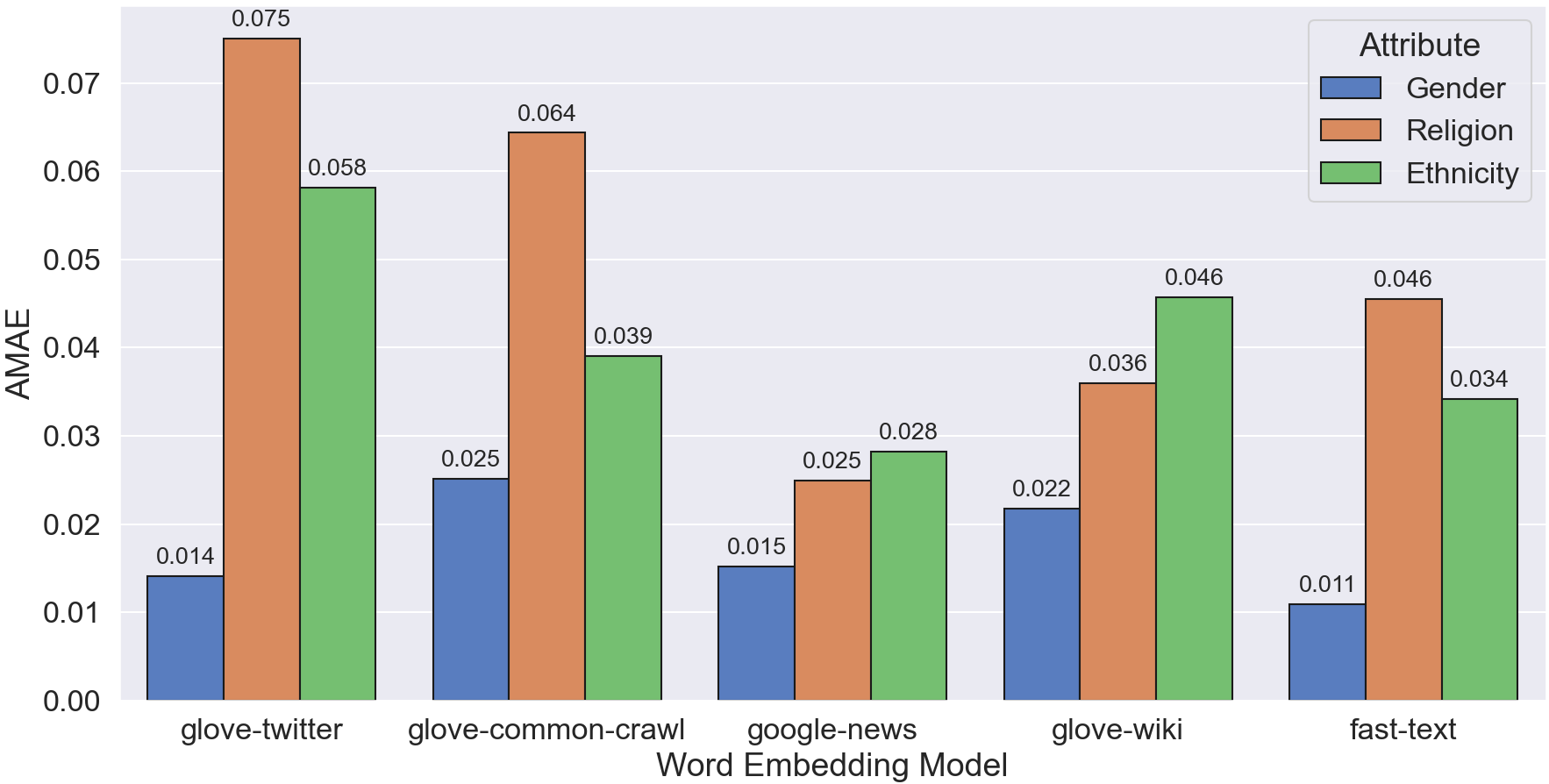}
  \caption{AMAE magnitude across different protected attributes and word embedding models}
  \label{fig:mae-across-encoders}
\end{figure}

\begin{figure}[h]
  \centering
  \includegraphics[width=\linewidth]{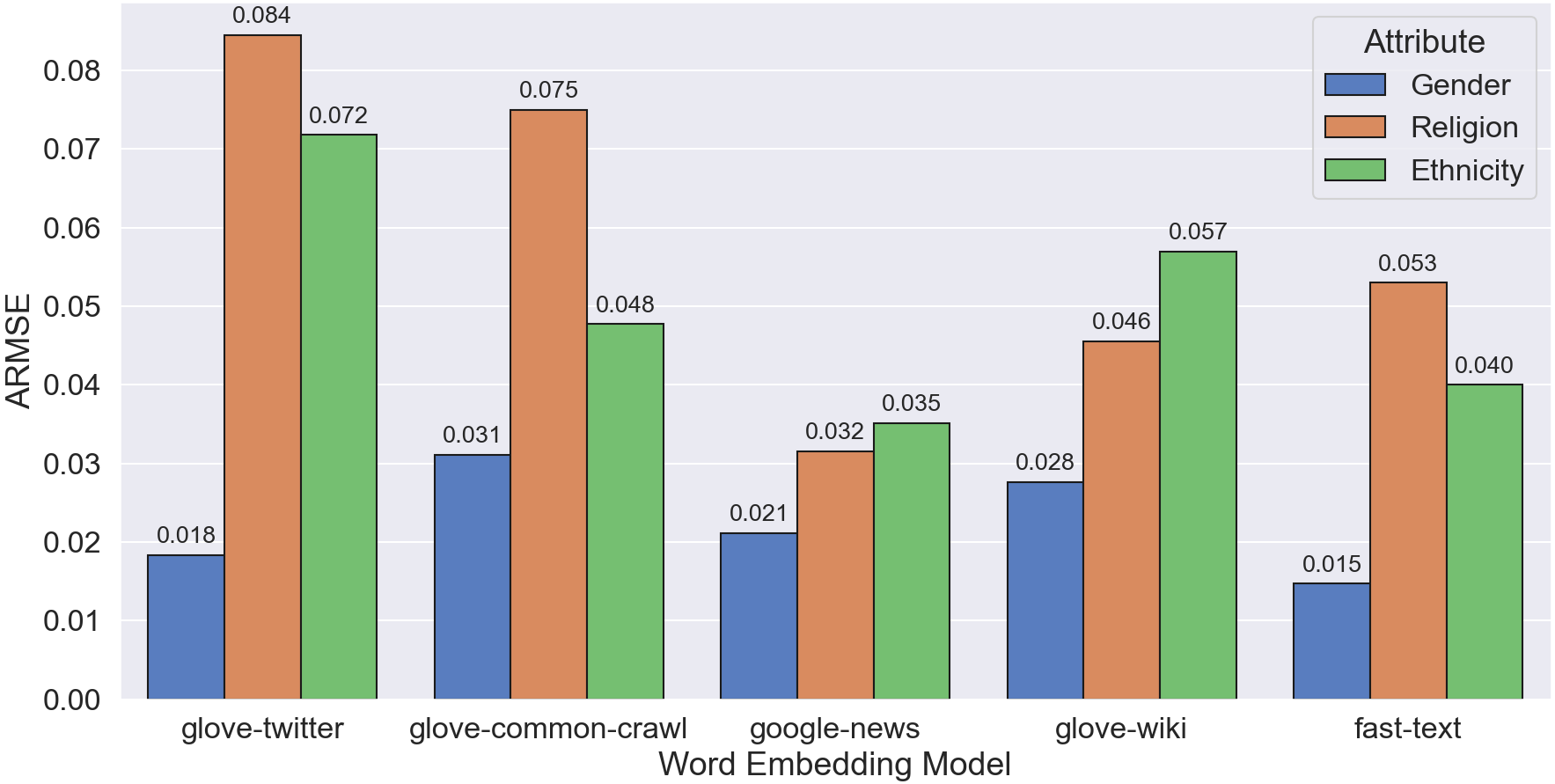}
  \caption{ARMSE magnitude across different protected attributes and word embedding models}
  \label{fig:rmse-across-encoders}
\end{figure}

\section{Classification bias metric definitions}
\label{appendix:bais-metric-def}
\emph{Equal Opportunity:} It measures the difference between true positive rates (proportion of samples correctly classified into positive class) for reference and protected groups.\\

\noindent \emph{Gini Equality:} It measures the difference between the Gini Inequality of benefits, which are defined as: Benefits = Prediction - Target + 1 \\

\noindent \emph{Normalized Treatment Equality:} It measures the difference between False Negative - False Positive ratio in reference and protected groups.\\

\noindent \emph{Overall Accuracy Equality:} It measures the difference between accuracy in reference and protected groups.\\

\noindent \emph{Positive Predictive Value:} It measures the difference between positive predictive values (proportion of positive samples among all samples classified as positive) for reference and protected groups.\\

\noindent \emph{Positive Class Balance:} It measures the difference between average predicted probability, given predicted class is positive, for reference and protected groups. \\

\noindent \emph{Statistical Parity:} It measures the difference between positive rates (proportion of samples classified into the positive class) for reference and protected groups.  \\

The choice of metrics depends on which type of error the user assigns more penalty to, and therefore wants to equalize across different groups. For example: Equal Opportunity means equalizing the True Positive Rate (TPR) across the different groups. If TPR is a criterion the user really cares about, then Equal Opportunity is a suitable fairness metric to consider. If the user cares more about the probability of getting the favourable result, the user should consider the Statistical Parity. Therefore, the selection of the metric depends on the user's needs and preferences. Automation of this process remains a challenge \cite{wachter2021fairness}.


\section{Classification bias supplementary data and results}
\label{appendix:classify-bias-supplements}
Tables \ref{tab:structured-data-entry}-\ref{tab:counterfactual-analysis-gender} provide supplementary results for classification bias assessment.

\begin{table}[ht]
  \caption{Examples of the structured dataset entries generated through protected attribute mining. Gender and religion are extracted using the look-up and NER based methods, respectively}
  \label{tab:structured-data-entry}
  \centering
  \begin{tabular}{p{0.53\linewidth} p{0.15\linewidth} p{0.20\linewidth}}
    \toprule
    \centering Comment & \centering Gender Extracted & \centering Religion Extracted \tabularnewline
    \midrule
    A visit to the DC Holocaust Museum revealed Hitler won by 43\% of the popular vote and 32\% of the seats. \underline{He} also used the Schutzstaffel (SS) to intimidate \underline{his} opponents, reminding one of the Antifa thugs we have today. Hitler also got the \underline{Muslims} on \underline{his} side. Uncanny. & \vfil{\hfil{Male}} & \vfil{\hfil{Islam}} \\
    \hline
    "Seton \underline{Catholic} where their own students talk trash about how low of a division there football team is in." & \vfil{\hfil{-}} & \vfil{\hfil{Christianity}} \\
    \hline
    How to develop their competitive skills and social skills in Home Schooling...as my child too complain abt bullying and unjust behaviour of teachers since \underline{she} is super active... & \vfil{\hfil{Female}} & \vfil{\hfil{-}} \\
  \bottomrule
  \end{tabular}
\end{table}

.

\begin{table}[ht]
  \caption{Swapped comments examples}
  \label{tab:swap-example}
  \begin{tabular}{c p{0.25\linewidth} ccc}
    \toprule
    Version & \centering Comment & Gender & Label & Prediction \\
    \midrule
    Original & This Jerry Lewis ripoff needs to just go away already. The guy is so good at acting like a fool because he is a fool. & Male & Hateful & 0.9811 \\ \hline
    Swapped & This Jerry Lewis ripoff needs to just go away already. The gal is so good at acting like a fool because she is a fool. & Female & Hateful & 0.9891 \\
   \bottomrule
  \end{tabular}
\end{table}

\begin{table}[ht]
  \caption{Synthetic counterfactual dataset examples for religion}
  \label{tab:counterfactual-example}
  \begin{tabular}{p{0.23\linewidth} p{0.23\linewidth} p{0.23\linewidth} p{0.12\linewidth}}
    \toprule
    \centering \vfil{Templates} & \centering Subgroup: Christianity & \centering Subgroup: Islam & \centering \vfil Label \tabularnewline
    \midrule
    I am a [Identity] person, ask me anything & I am a Christian person, ask me anything & I am a Muslim person, ask me anything & Not-hateful \\ \hline
    [Identity] people are gross and universally terrible & Christian people are gross and universally terrible & Muslim people are gross and universally terrible & \hfil Hateful \\
   \bottomrule
  \end{tabular}
\end{table}

\begin{table}[ht]
  \caption{Prediction probability statistics for counterfactual comments with religion references}
  \label{tab:counterfactual-analysis-religion}
  \begin{tabular}{p{0.28\linewidth} p{0.28\linewidth} p{0.28\linewidth}}
    \toprule
    \centering Actual Comment Type & \centering Subgroup & \centering Avg. Predicted Probability \tabularnewline
    \midrule
    \hfil Not-hateful & \hfil Islam & \hfil 15.6\% hateful \\
    \hfil Not-hateful & \hfil Christianity & \hfil 11.2\% hateful \\
    \hfil Hateful & \hfil Islam & \hfil 67.9\% hateful \\
    \hfil Hateful & \hfil Christianity & \hfil 63.5\% hateful \\
  \bottomrule
  \end{tabular}
\end{table}

\begin{table}[ht]
  \caption{Prediction probability statistics for counterfactual comments with gender references}
  \label{tab:counterfactual-analysis-gender}
  \begin{tabular}{p{0.28\linewidth} p{0.28\linewidth} p{0.28\linewidth}}
    \toprule
    \centering Actual Comment Type & \centering Subgroup & \centering Avg. Predicted Probability \tabularnewline
    \midrule
    \hfil Not-hateful & \hfil Male & \hfil 12.9\% hateful \\
    \hfil Not-hateful & \hfil Female & \hfil 13.4\% hateful \\
    \hfil Hateful & \hfil Male & \hfil 57.4\% hateful \\
    \hfil Hateful & \hfil Female & \hfil 57.2\% hateful \\
  \bottomrule
  \end{tabular}
\end{table}

\section{Global interpretability results}
\label{appendix:global-interpret-result}
\begin{figure}[h]
  \centering
  \includegraphics[width=\linewidth]{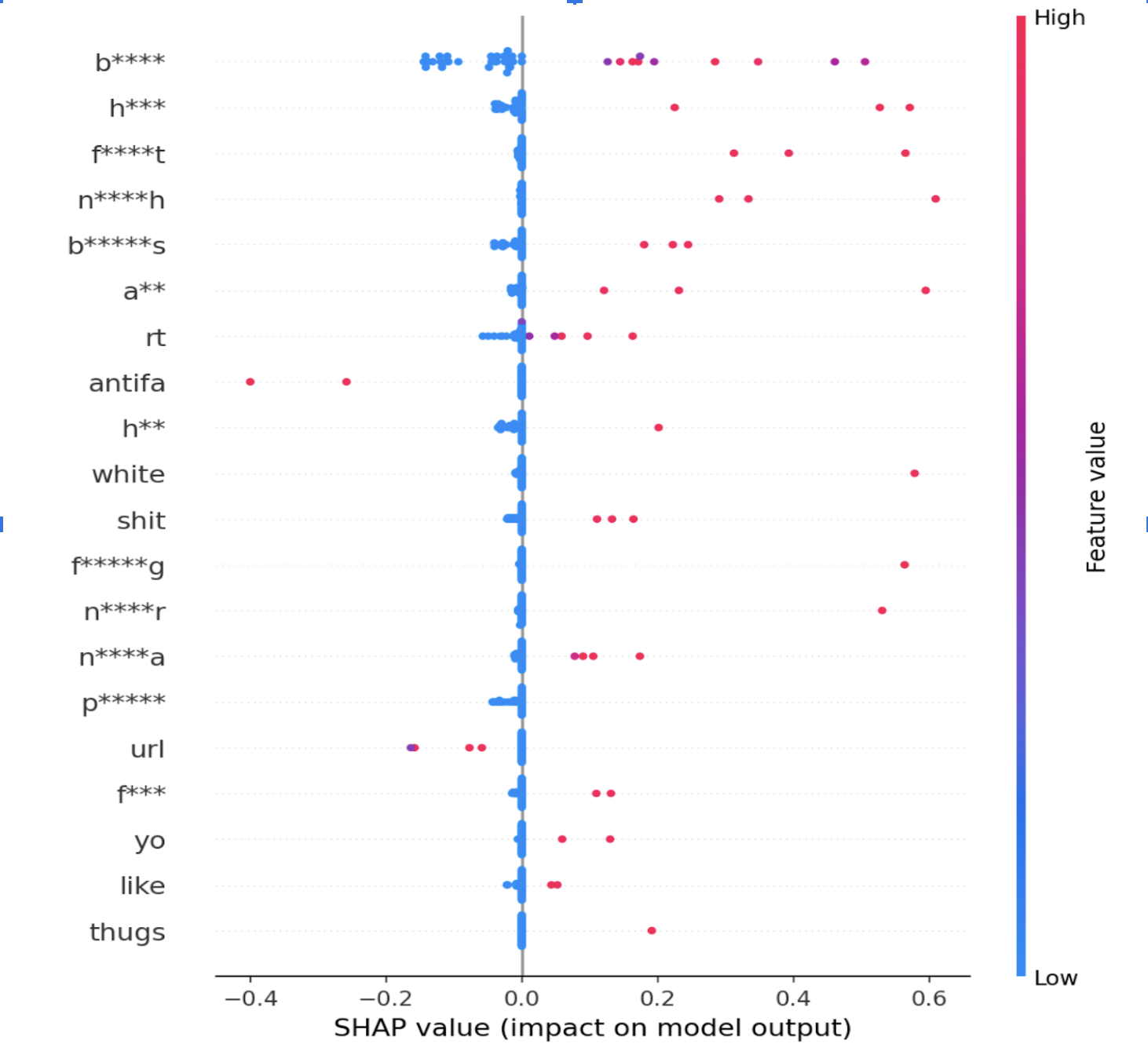}
  \caption{Global interpretability of predictions using SHAP}
  \label{fig:shap-example}
\end{figure}

The global interpretability of predictions using SHAP provides a holistic view of how a given token impacts the overall working of the model. The results are illustrated in Figure \ref{fig:shap-example}. The various dots in this plot signify the presence of a token across different sets of comments. The vertical location shows what token is being depicted, the color shows whether that token’s importance was high or low for that observation, and the horizontal location shows whether the effect of that value caused a higher or lower prediction (model’s output probability). We generated this view for the hate-comments in particular, and discovered that profane language or cuss words turned out to be significant and positively impacting the model - thereby driving it to classify these comments as “hateful”.

\section{Other ALTAI Requirements}
\label{appendix:environ-impact-assess}
ALTAI Requirement \#6 focuses on societal and environmental well-being and considers the environmental impact of AI systems. There exist different methods and mechanisms that can be used to evaluate the associated environmental impacts. Our process uses the Machine Learning Emissions Calculator tool \cite{lacoste2019quantifying} to estimate the amount of carbon emissions produced by training the classifier under assessment here. This can be measured in Carbon dioxide equivalent (CO2eq) units, which is “a metric measure used to compare the emissions from various greenhouse gases on the basis of their global-warming potential, by converting amounts of other gases to the equivalent amount of carbon dioxide with the same global warming potential.” \cite{eurostat2017glossary}. The model assessed in this study has been trained on CPUs locally and its carbon footprint was considered to be negligible.

There are several additional ethical AI requirements that have not been been applicable or covered in this paper. The model under assessment here has been trained offline on reliable datasets. But resilience of the model to cyber-attacks, such as data poisoning and model evasion, is a critical dimension if the model is trained online and using comments labels through crowdsourcing (ALTAI Requirement \#2). The model here used public datasets. But in the event of use of or application to private data, the assessment should also evaluate privacy dimensions (ALTAI Requirement \#3). This also includes data governance on both a macro and a micro level, with the focus areas of data availability, sharing, usability, consistency, integrity.

The assessment may also account for the interests of stakeholders and how much each requirement is important to each. The idea of relevance matrix proposed by \cite{brown2021algorithm} can be used to connect them with the models ethical performance. The assessment output can also inform AI impact assessment \cite{kazim2021ai}, which are usually first party created, and not necessarily covered in second/third party reviews.

\end{document}